\documentclass[journal]{IEEEtran}

\usepackage{xcolor,soul,framed, eufrak, cite} 

\colorlet{shadecolor}{yellow}
\usepackage[pdftex]{graphicx}
\usepackage{tabularx}
\usepackage{etoolbox}
\usepackage{booktabs}
\usepackage{multirow}
\usepackage{caption}
\usepackage{subcaption}

\def\tablename{Table}
\graphicspath{{../pdf/}{../jpeg/}}
\DeclareGraphicsExtensions{.pdf,.jpeg,.png}

\usepackage[cmex10]{amsmath}
\usepackage{amsfonts, amsmath}
\usepackage{mathtools}
\usepackage{array}
\usepackage{mdwmath}
\usepackage{mdwtab}
\usepackage{eqparbox}
\usepackage{url}
\usepackage{stmaryrd}
\newcounter{assume}
\newcounter{propose}
\newcounter{thms}
\newcounter{define}
\newtheorem{assumption}[assume]{Assumption}
\newtheorem{proposition}[propose]{Proposition}
\newtheorem{definition}[define]{Definition}
\newtheorem{theorem}[thms]{Theorem}
\newtheorem{remark}{Remark}

\newcommand{\cpd}[3]{\llbracket #1, #2, #3 \rrbracket}
\newcommand{\A}{{\bf A}}
\newcommand{\B}{{\bf B}}
\newcommand{\C}{{\bf C}}
\newcommand{\Ten}[1]{\mathcal{#1}}
\newcommand{\floor}[1]{\lfloor \log_2{#1}\rfloor}

\newcommand{\bc}{{\bf c}}
\newcommand{\ba}{{\bf a}}
\newcommand{\bb}{{\bf b}}

\begin{document}
\bstctlcite{IEEEexample:BSTcontrol}
    \title{Model-Free State Estimation Using Low-Rank Canonical Polyadic Decomposition}
     \author{Ahmed S. Zamzam,~\IEEEmembership{Member,~IEEE,}
      Yajing Liu,~\IEEEmembership{Member,~IEEE,}
      Andrey Bernstein,~\IEEEmembership{Member,~IEEE,}
  \thanks{The authors are with the National Renewable Energy Laboratory, Golden, CO, USA 80401 (e-mails: \texttt{\{ahmed.zamzam, yajing.liu, andrey.bernstein\}@nrel.gov}).\newline
   This work was authored by the National Renewable Energy Laboratory, operated by Alliance for Sustainable Energy, LLC, for the U.S. Department of Energy (DOE) under Contract No. DE-AC36-08GO28308. Thir work was supported by the Laboratory Directed Research and Development Program at the National
Renewable Energy Laboratory.  The views expressed in the article do not necessarily represent the views of the DOE or the U.S. Government. The U.S. Government retains and the publisher, by accepting the article for publication, acknowledges that the U.S. Government retains a nonexclusive, paid-up, irrevocable, worldwide license to publish or reproduce the published form of this work, or allow others to do so, for U.S. Government purposes.}
  }  

\maketitle

\begin{abstract}
As electric grids experience high penetration levels of renewable generation, fundamental changes are required to address real-time situational awareness. This paper uses unique traits of tensors to devise a model-free situational awareness and energy forecasting framework for distribution networks. This work formulates the state of the network at multiple time instants as a three-way tensor; hence, recovering full state information of the network is tantamount to estimating all the values of the tensor. Given measurements received from $\mu$phasor measurement units and/or smart meters, the recovery of unobserved quantities is carried out using the low-rank canonical polyadic decomposition of the state tensor---that is, the state estimation task is posed as a tensor imputation problem utilizing observed patterns in measured quantities. Two structured sampling schemes are considered: slab sampling and fiber sampling. For both schemes, we present sufficient conditions on the number of sampled slabs and fibers that guarantee identifiability of the factors of the state tensor. Numerical results demonstrate the ability of the proposed framework to achieve high estimation accuracy in multiple sampling scenarios.
\end{abstract}


\IEEEpeerreviewmaketitle

\section{Introduction}
\IEEEPARstart{T}{he} rapid integration of renewable generation sources and responsive electric loads increases the complexity of the operation of electric power distribution networks and calls for distribution system control and imputation techniques, such as state estimation. The state estimation task is usually formulated to infer the state of the system from physical measurements.
Traditional state estimation methods for transmission systems typically rely on: (i)
the availability of accurate measurements and full observability (i.e., the number of measurements is at least equal to the number of unknown states); and (ii) the knowledge of the network model (i.e., bus admittance matrix). In such scenarios, state estimation methods are typically formulated as variants of  least-squares estimators~\cite{baran1994state, kekatos2012distributed}.

In distribution systems, however, these assumptions are not realistic: measurements are scarce compared to the number of unknowns, and models are hard to come by (or, at best, inaccurate) because of aging infrastructure or unrecorded topological changes. 
Many approaches were proposed in the literature to address the challenges of achieving reliable estimation in situations with limited observability \cite{donti2019matrix, Sagan2019, zamzam2019data, wang2019distribution, Cav_GS19, zhou2019gradient}. A standard approach to alleviating this challenge is to leverage  pseudo-measurements in the form of load or energy forecast~\cite{manitsas2012distribution}. In addition, Bayesian estimation approaches were also proposed when historical data are available to learn the underlying mapping from measurements to states~\cite{mestav2019bayesian, zamzam2019physics, ostrometzky2019physics, zhang2019real}; however, these methods require either accurate knowledge of the network model or full observability of all quantities at all buses to generate training samples.
Obtaining an accurate model of large distribution systems is a cumbersome task that hinders the applicability of classic state estimation approaches; thus, the need for model-free state estimation approaches becomes prominent. 

In this work, we use some notion of regularity in the state information over time to introduce a parsimonious model that allows imputing unavailable information or states in the challenging scenario of low observability and unavailability of the system model. The use of data collected at different time instances---and hence different operational situations---allows for exploiting observed patterns to build a more comprehensive picture of the network operation.

Tensors are a natural generalization of matrices where each tensor element is indexed by three or more indices. Tensor factorization methods have been widely used for data mining and analysis tasks~\cite{kolda2005higher, karatzoglou2010multiverse}; see~\cite{Sidiropoulos2017} for a comprehensive overview of tensor algebra and applications. In this work, the state information of the network is stacked in a tensor representing the state at multiple time instants. The modes of the state tensor represent buses, measurements, and time instances. Often, the tensor cannot be fully observable; hence, the task of state estimation or forecasting is formulated as a tensor completion. To impute unobserved (missing) elements in the state tensor, we use the canonical polyadic decomposition (CPD) of the state tensor. Contrary to matrix completion methods, the CPD factors can be essentially unique under mild conditions~\cite{Sidiropoulos2017}, which lead to recovery guarantees of the ground-truth factors as well as the ability to impute unobserved (missing) values.

Tensors have been used as a tool to analyze power systems data~\cite{song-2017} where the  tensor model was learned from historical data to build a power flow model. Then this model was used to forecast network states. This method requires training data that comprise detailed information about voltages and current injections throughout the network, which might not be easy to obtain. More recently, a model-based state estimation method was formulated using a tensor decomposition model~\cite{madbhavi2020tensor}. A nuclear-norm minimization optimization framework was used to estimate the future state of distribution systems where the network model is assumed to be known a priori.

In this work, we formulate the state estimation as a tensor completion from structured sampling schemes. Evidently, the constructed state tensors have a low-rank structure that can be used to impute unobserved elements. Based on the identifiability conditions of sampled tensors, we can derive theoretical guarantees on sufficient conditions for recovering the latent factors of the state tensor. Although the proposed framework can be generalized to other unstructured or random sampling schemes, this work considers slab and fiber sampling paradigms motivated by recent interesting results in~\cite{kanatsoulis2018hyperspectral, Kanatsoulis2020}. The derived sufficient conditions can also be useful for measuring devices placements. We assess the ability of the proposed approach to impute unobserved data in the state tensor on the $123$-bus distribution feeder. The proposed model-free approach achieves accurate estimation performance under both the slab and fiber sampling schemes.

Throughout this paper, matrices (vectors) are denoted by upper- (lower-) case boldface letters. Tensors and sets are denoted by calligraphic letters, and it will be clear from the context. Symbol $\mathfrak{Re}\{.\}$ ($\mathfrak{Im}\{.\}$) takes the real (imaginary) part of a complex-valued object, and $|.|$ denotes the cardinality of sets.

This paper is organized as follows. Tensor preliminaries and CPD definitions are outlined in Section II. We present the problem statement and the proposed optimization formulations in Section III. Sufficient conditions for recovering latent CPD factors are presented in Section IV, and simulation results are presented in Section V. Finally, Section VI concludes the work and outlines future research directions.

\section{Tensor Preliminaries}


\subsection{Canonical Polyadic Decomposition}
\emph{Tensors} of order higher than two are arrays indexed by three or more indices. Consider a  third-order tensor $\Ten{X}\in\mathbb{R}^{I\times J\times K}$ with elements $\Ten{X}(i,j,k)$.
A tensor $\Ten{X}\in\mathbb{R}^{I\times J\times K}$ is \emph{rank one} if it can be written as the outer product of three vectors, i.e.:
\[\mathcal{X} = {\bf a} \circ {\bf b}\circ {\bf c},\]
where ${\bf a}\in\mathbb{R}^I$, ${\bf b}\in\mathbb{R}^J$, ${\bf c}\in\mathbb{R}^K$, and the symbol ``$\circ$'' represents the vector outer product.

A \emph{polyadic decomposition} represents a tensor $\mathcal{X}\in\mathbb{R}^{I\times J\times K}$ as a linear combination of rank one tensors in the form: 
\begin{equation}
    \label{eq:CPD}
    \mathcal{X} = \sum\limits_{f=1}^F {\bf a}_f\circ {\bf b}_f \circ {\bf c}_f,
\end{equation}
where $F$ is a positive integer, and ${\bf a}_f \in\mathbb{R}^I, {\bf b}_f\in\mathbb{R}^J, {\bf c}_f\in\mathbb{R}^K$ for $f=1,\ldots, F$.
The smallest value of $F$ for which \eqref{eq:CPD} holds is defined as the \emph{tensor rank} or the \emph{PD rank}, and the minimum rank polyadic decomposition is called the \emph{canonical polyadic decomposition} (CPD). 
Let $\A =[\ba_1,\ldots, \ba_F]\in\mathbb{R}^{I\times F}$, $\B =[\bb_1,\ldots, \bb_F]\in\mathbb{R}^{J\times F}$, and $\C =[\bc_1,\ldots, \bc_F]\in\mathbb{R}^{K\times F}$. 
Then we use the notation $\Ten{X} = \cpd{\A}{\B}{\C}$ to denote \eqref{eq:CPD}.
\subsection{Identifiability}
The CPD of a tensor $\Ten{X}$ is \emph{essentially unique} under mild conditions. The definition of essential uniqueness is presented as follows.
\begin{definition}[Essential uniqueness~\cite{Sidiropoulos2017}]
The CPD of a tensor $\Ten{X}\in\mathbb{R}^{I\times J\times K}$,  $\Ten{X} = \cpd{\A}{\B}{\C}$ is \emph{essentially unique} if $\A,\B,\C$ are unique up to unavoidable scaling and permutation, i.e., if $\Ten{X} = \cpd{\bar{\A}}{\bar{\B}}{\bar{\C}}$ for some $\bar{\A}\in\mathbb{R}^{I\times F}$, $\bar{\B}\in\mathbb{R}^{J\times F},$ and $\bar{\C}\in\mathbb{R}^{K\times F}$, then there exists a permutation matrix $\bf{\Pi}$ and diagonal scaling matrices $\bf{\Lambda_1}, \bf{\Lambda_2}, \bf{\Lambda_3}$ such that:
\[\bar{\A} = \A\bf{\Pi} \bf{\Lambda_1}, \bar{\B} = \B\Pi \bf{\Lambda_2}, \bar{\C} = \C\Pi \bf{\Lambda_3}, \bf{\Lambda_1}\bf{\Lambda_2}\bf{\Lambda_3} = I.\]
\end{definition}

A remarkable property of tensors is that the CPD can be unique even for rank $F$ much higher than any dimension of the tensor, i.e., $F \gg I, J, K$. A generic uniqueness condition of the CPD  is given as follows.
\begin{theorem}[\cite{Chiantini2012}]
Let $\Ten{X} = \cpd{\A}{\B}{\C}$ with $\A\in\mathbb{R}^{I\times F}$, $\B\in\mathbb{R}^{J\times F}$, and $\C\in\mathbb{R}^{K\times F}$. Assume that $\A, \B,$ and $\C$ are drawn from
some joint absolutely continuous distribution and $I\geq J\geq K$ without loss of generality. If: 
\begin{equation}
    \label{eq:indetfiability}
    F\leq 2^{\floor{J}+\floor{K}-2},
\end{equation}
then the decomposition of $\Ten{X}$ in terms of $\A,\B,\C$ is
essentially unique, almost surely.
\end{theorem}

Other uniqueness conditions are based on the Kruskal rank of matrices $\A, \B, $ and $\C$; see \cite{Kruskal1977, Sidiropoulos2017} for  details.
\subsection{Sampling Schemes}
In this work, we are motivated by \cite{Kanatsoulis2020}, which states that any third-order tensor of a reasonably low rank can be recovered with sufficient samples using some structured sampling schemes. We consider two classes of sampling schemes: \emph{slab sampling} and \emph{fiber sampling}. The related definitions and notation follow.

\emph{A slab} is a two-dimensional section of a tensor---i.e., a matrix---defined by fixing all but two indices \cite{Kolda2009}. Using the MATLAB indexing format, a tensor $\Ten{X}\in\mathbb{R}^{I\times J \times K}$ has three types of slabs: \emph{horizontal} $\Ten{X}(i,:,:)$, \emph{vertical} $\Ten{X}(:,j,:)$, and \emph{frontal} $\Ten{X}(:,:,k)$. For horizontal slab sampling, we collect the sampled horizontal slabs with indices collected in $\mathcal{I}_h \subseteq \{1, 2, \ldots, I\}$; hence, a sampled subtensor $\Ten{Y}_h \in \mathbb{R}^{|\mathcal{I}_h| \times J \times K}$ is constructed. Similarly, for vertical and frontal slabs, the sampled slabs are collected in tensors $\Ten{Y}_v \in \mathbb{R}^{I \times |\mathcal{J}_v|\times K}$ and $\Ten{Y}_f \in \mathbb{R}^{I \times J\times |\mathcal{K}_f|}$, respectively, where $\mathcal{J}_v \subseteq \{1, 2, \ldots, J\}$ and $\mathcal{K}_f \subseteq \{1, 2, \ldots, K\}$.

\emph{A mode} is a one-dimensional section of a tensor---i.e., a vector---defined by fixing every index but one. A tensor  $\Ten{X}\in\mathbb{R}^{I\times J \times K}$ has three types of modes:
\emph{columns} $\Ten{X}(:,j,k)$, \emph{rows} $\Ten{X}(i,:,k)$, and \emph{fibers} $\Ten{X}(i,j,:)$. In this paper, we consider only fiber sampling. We use $d$ to denote the pattern index of the fiber-sampling scheme. Every sampled fiber $\Ten{X}(i,j,:)$ belongs to a pattern with a subset of rows $\mathcal{S}_r^{(d)}\subseteq \{1,\ldots, I\}$ and  columns $\mathcal{S}_c^{(d)}\subseteq \{1,\ldots, J\}$. And we use $\Ten{Y}_d\in\mathbb{R}^{|\mathcal{S}_c^d|\times \mathcal|{S}_r^d| \times K}$ to denote the sampled subtensor using pattern $d$.


\section{Problem Statement}
In distribution systems, the state estimation problem aims to estimate the state of the feeder from available measurements. When an accurate model of the network is available, the voltage phasors at all buses/phases in the network are enough to recover any other quantity, e.g., power consumption/injection and current flows~\cite{wang2018power}. However, such an accurate model of the feeder and the line parameters is often not at the system operator's disposal. This is because of the change in line parameters resulting from aging and/or unrecorded topology changes. Therefore, we define the state of the network as the collection of active and reactive power injections as well as the voltage phasors at all buses/phases.

We construct a three-way tensor $\Ten{X}$ that collects nodal measured or forecasted quantities for multiple time instants. The tensor modes are \textsc{PHASE $\times$ MEASUREMENT $\times$ TIME}\footnote{By construction, multiphase nodes are represented using multiple elements in the tensor.}. That is, the measurement $j$ taken at phase $i$ at time instant $k$ is placed in the $(i, j, k)$-th element of the tensor $\Ten{X}$. 
The considered nodal measurements are voltage phasor real and imaginary parts ($\mathfrak{Re}\{v_i\}$, $\mathfrak{Im}\{ v_i\}$), nodal voltage magnitude ($|v_i|$), and active and reactive power injections ($p_i$, $q_i$).

The CPD factors of  tensor $\Ten{X}$ are denoted by $\A$, $\B$, and $\C$. Therefore, the elements of  tensor $\Ten{X}$ can be written as:
\begin{align}
\Ten{X}(i, j, k) = \sum_{f=1}^{F} \A(i, f) \B(j, f) \C(k, f),
\end{align}
where $F$ denotes the rank of the tensor. In practice, the rank of the state tensor is low, which will be demonstrated on real data in the simulations section.
\begin{figure}
	\centering
	\begin{subfigure}[b]{0.217\textwidth}
		\centering
		\includegraphics[width=\textwidth]{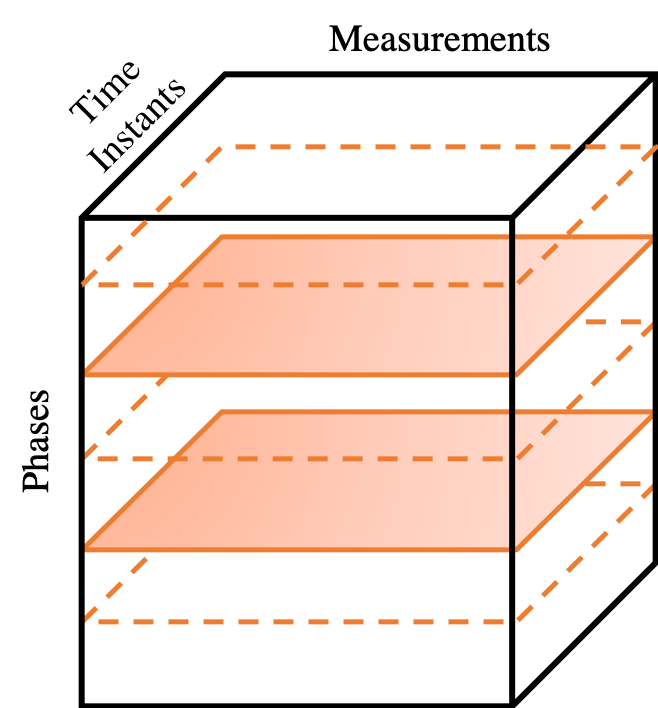}
		\caption{Horizontal slab sampling}
		\label{fig:horizontal-slab}
	\end{subfigure}
	\hspace{10pt}
	\begin{subfigure}[b]{0.22\textwidth}
		\centering
		\includegraphics[width=\textwidth]{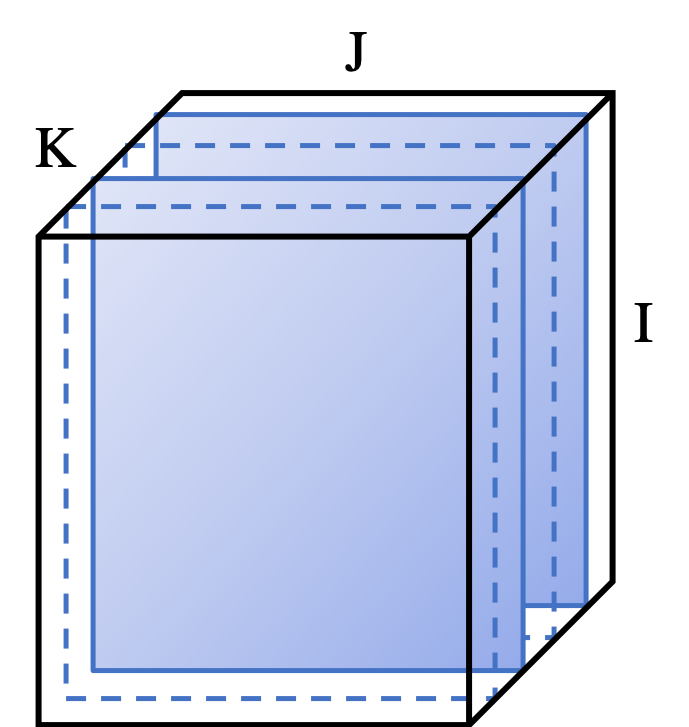}
		\caption{Frontal slab sampling}
		\label{fig:frontal-slab}
	\end{subfigure}
	\caption{Two slab sampling schemes. Dashed slabs refer to unobserved slabs, whereas the sampled entries are colored.}
	\label{fig:slab-sampling}
\end{figure}

\subsection{Slab Sampling}
The first imputation problem we tackle stems from a slab-sampling scenario. By our construction of the state tensor $\Ten{X}$, a sampled horizontal slab collects all types of measurements at all time steps for a specific phase (bus). The phases (buses) with such information are usually equipped with phasor measurement units where phasor measurements are collected with high frequency. 
In this sampling scheme, the phases (buses) with all measurements available at all time slots are collected in the set $\mathcal{I}_h$, with cardinality $I_h$. In addition, frontal slab sampling amounts to knowing the state of the whole network at a specific time step, i.e., all measurement types at all phases. This type of data is a bit unrealistic to obtain directly, but it can be obtained indirectly through load forecasting or state estimation algorithms that can be used when enough measurements are available to make the network observable at only a few time instants. The sampled frontal slab indices are collected in the set $\mathcal{K}_f$, where $K_f := |\mathcal{K}_f|$.

The model-free state estimation problem aims at imputing the tensor $\Ten{X}$ using the sampled frontal and horizontal slabs. We introduce the selection tensor $\Ten{M}_s$, which has ones at the locations observed (sampled) and zero otherwise. In other words, the $(i, j, k)$-th element of the tensor $\Ten{M}_s$ is given by:
\begin{align*}
\Ten{M}_s(i, j, k) =  1 \ \text{if $i \in \mathcal{I}_h$} \ \text{or $k \in \mathcal{K}_f$}; \              0 \ \text{otherwise}.
\end{align*}
This sampling tensor will be used to formulate the factorization optimization problem to recover the latent factors $\A$, $\B$, and $\C$. The next section presents sufficient conditions on ${I}_h$ and ${K}_f$ to guarantee recovery of latent factors.

\begin{figure}
	\centering
	\includegraphics[width=0.22\textwidth]{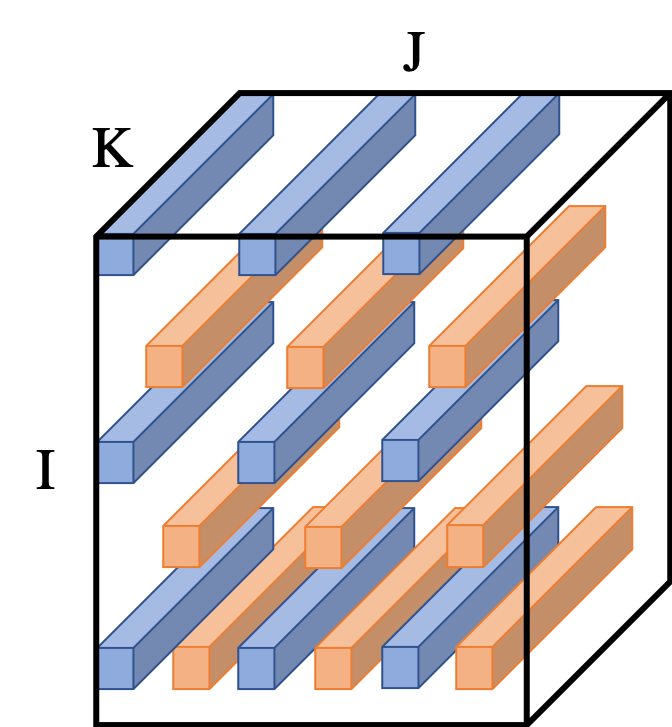}
	\label{fig:fiber-sampling}
	\caption{Fiber-sampling example with two sampling patterns}\vspace{-20pt}
\end{figure}

\subsection{Fiber Sampling}
We also consider a structured fiber-sampling scheme. In this scheme, each fiber is indexed by two indices, where the first index determines the specific phase, and the second indicates the type of measurement; hence, each fiber collects a specific measurement at a specific phase for all time instances. The structure of each sampling pattern consists of selecting specific rows and columns where the fibers are sampled. This fiber, for example, can collect the voltage magnitude at a specific bus for all time instances that occur when a voltage magnitude measuring device is installed. For a set of rows $\mathcal{S}_r^{(d)} \subseteq \{1, 2, \ldots, I\}$ and a set of columns $\mathcal{S}_c^{(d)} \subseteq \{1, 2, \ldots, J\}$, we say that the fibers with indices $(i, j)$ such that $i \in \mathcal{S}_r^{(d)}$ and $j\in\mathcal{S}_c^{(d)}$ are sampled in the $d$-th sampling scheme. 

We consider the case when we have two patterns of fiber sampling, i.e., $d\in \{1, 2\}$. We define a tensor $\Ten{M}_f$ that has the same dimension as $\Ten{X}$ and has ones only at the observed (sampled) locations; hence, the elements of $\Ten{M}_f$ are given by:
\begin{align*}
\Ten{M}_f(i, j, k) = \begin{cases} 1 &\quad \text{if $i \in \mathcal{S}_r^{(1)}$ and $j \in \mathcal{S}_c^{(1)}$}, \\
1 &\quad \text{if $i \in \mathcal{S}_r^{(2)}$ and $j \in \mathcal{S}_c^{(2)}$}, \\
0 &\quad \text{otherwise.}
\end{cases}
\end{align*}
This binary will be used to formulate an optimization problem to recover the latent factors. In the next section, we present sufficient conditions on $\mathcal{S}_r^{(d)}$ and $\mathcal{S}_c^{(d)}$ such that the ground-truth factors $\A$, $\B$, and $\C$ can be recovered.

In general, the samples available from the state tensor $\Ten{X}$ might not have a specific structure, such as the slab or fiber sampling. In such a case, a tensor that has ones only at the observed locations and zeros otherwise can be used. However, deriving sufficient conditions for recovering the factors becomes more challenging in this case, and it is left for future research.

\subsection{Imputation Optimization Problem}
Using the sampling tensor $\Ten{M}_s$ or $\Ten{M}_f$, the optimization problem that imputes the unobserved values of $\Ten{X}$ can be formulated as:
\begin{align}\label{eq:impute-opt}
\min_{\A, \B, \C} \| \Ten{M} * \big(\Ten{X} - {\cpd{\A}{\B}{\C}} \big)\|_F^2,
\end{align}
where $\Ten{M}$ is $\Ten{M}_s$ or $\Ten{M}_f$ in the slab or fiber sampling cases, respectively.

\begin{remark}
	The optimization problem~\eqref{eq:impute-opt} can be tackled directly using the Tensorlab MATLAB toolbox~\cite{tensorlab}. In all our simulations, the Gauss-Newton solver within the toolbox was used.
\end{remark}

\section{Sampling Requirements}
\label{sec:samplingrequirements}
Following the intuitive representation of the state estimation problem as a tensor completion problem in the previous section, the goal of this section is to provide theoretical bound on the minimum requirements to guarantee identifiability of low-rank factors from available data. First, we introduce the following assumption which will be verified on real data in Section \ref{sec:num}. 

\begin{assumption}
Assume that the state tensor $\Ten{X}$ adheres to a CPD representation as in~\eqref{eq:CPD} with rank $F$. Also, assume that the rank $F$ satisfies the identifiabiablity condition~\eqref{eq:indetfiability}.
\end{assumption}

The previous assumption states that the full state tensor has a relatively low-rank structure, and thus, using the CPD model is feasible to impute missing values in $\Ten{X}$. 

\subsection{Slab sampling}
\label{subsec:slabsampling}
\begin{proposition}
\label{prop:slabsampling}
In the case of slab sampling, if the factors of state tensor $\Ten{X}$ are drawn from an absolutely continuous distribution and the numbers of sampled phases ($I_h$) and sampled time instances ($K_f$) satisfy one of the following conditions:
\begin{enumerate}
   \item $
        \min \big\{\floor{I_h} + \floor{J}, \floor{J} + \floor{K},$ \\\phantom{x}\hspace{10pt} $\quad  \floor{I_h} + \floor{K}, \log_2{(4 J K_f)} \big\} \geq \log_2{ (4 F)}
    $,
    \item $
        \min \big\{\floor{I} + \floor{J}, \floor{J} + \floor{K_f},$ \\\phantom{x}\hspace{10pt} $\quad  \floor{I} + \floor{K_f}, \log_2{(4 I_h J)} \big\} \geq \log_2{ (4 F)}
    $,
\end{enumerate}
then the ground truth factors can be recovered almost surely from the observed elements.
\end{proposition}
This proposition can be proven using the same proof technique used in~\cite[Theorem 3]{Kanatsoulis2020}. In the proof, we construct subtensors using the sampled slabs. Note that each of the subtensors shares two latent factors with the state tensor $\Ten{X}$, and hence, two of the latent factors can be recovered if the one of the subtensors satisfies the identifiability conditions in Theorem 1. The conditions 1) and 2) in the proposition guarantee that one of the two subtensors have indentifiable latent factors. For instance, if the sampling scheme achieves the first condition, then having the minimum of the first three terms less than $\log_2{(4F)}$ guarantees that the latent factors $\B$, and $\C$ can be identified from the subtensor containing the sampled horizontal slabs. In addition, having $J K_f \geq F$ guarantees that by matricizing the other subtensor, the resulting matrix will be full rank~\cite{jiang2001almost}, and thus, the latent factor $\A$ can be recovered from this matrix. Detailed proof is available in~\cite{Kanatsoulis2020}, and it is omitted here for brevity.

\subsection{Fiber sampling}
\label{subsec:fibersampling}
\begin{proposition}
\label{prop:fibersampling}
In the fiber sampling scheme, if the factors of the state tensor $\Ten{X}$ are drawn from an absolutely continuous distribution and the number of sampled fibers in sampled pattern satisfy the following conditions:
\begin{enumerate}
    \item $|\mathcal{S}_r^{(d)}|, |\mathcal{S}_c^{(d)}|\geq 2$,
    \item $\mathcal{S}_r^{(1)} \bigcup \mathcal{S}_r^{(2)} = \{1, 2, \ldots, I\}$,
    \item $\mathcal{S}_c^{(1)} \bigcup \mathcal{S}_c^{(2)} = \{1, 2, \ldots, J\}$,
    \item $\big(\mathcal{S}_r^{(1)} \cap \mathcal{S}_r^{(2)}\big)\ \bigcup\ \big(\mathcal{S}_c^{(1)} \cap \mathcal{S}_c^{(2)}\big) \neq \emptyset$,
    \item $\min_{d\in \{1, 2\} } \big\{ \floor{|\mathcal{S}_r^{(d)}|} + \floor{|\mathcal{S}_c^{(d)}|}, \floor{|\mathcal{S}_r^{(d)}|} + $\\ \phantom{x}\hspace{10pt} $\quad\floor{K}, \floor{|\mathcal{S}_c^{(d)}|} + \floor{K}\big\} \geq \log_2(4 F)$,
\end{enumerate}
then the ground truth factors can be recovered almost surely from the observed elements.
\end{proposition}

This proposition can be proven following the proof technique in~\cite[Theorem 4]{Kanatsoulis2020}. First, we construct two subtensors that collect the sampled entries of the state tensor $\Ten{X}$. The dimensions of the subtensors are $\Ten{T}_d \in \mathbb{R}^{|\mathcal{S}_r^{(d)}| \times |\mathcal{S}_c^{(d)}| \times {K}}$ for $d \in \{1, 2\}$. Then, the conditions (1) and (5) guarantee that the indentifiability condition in Theorem 1 is satisfied for both $\Ten{T}_1$ and $\Ten{T}_2$. This means that the factor $\C$ can be recovered from the CPD of either subtensor. The conditions (2) and (3) guarantee that all rows of the latent factors $\A$ and $\B$ are sampled in the factors of the subtensors $\Ten{T}_1$ and $\Ten{T}_2$. Finally, condition (4) guarantees that the rotation ambiguities can be resolved given that the factors $\A$ and $\B$ are drawn from absolutely continuous distribution~\cite{jiang2001almost}. Thus, the latent factors can be identified if the aforementioned conditions are satisfied. For detailed proof of the generic fiber sampling scheme, we refer the reader to~\cite{Kanatsoulis2020}.

\section{Numerical Results} \label{sec:num}
We implement the slab- and fiber-sampling schemes, which lead to solving the optimization problem~\eqref{eq:impute-opt} for the IEEE 123-bus system using the \textsc{sdf-nls} solver from~\cite{tensorlab}. The IEEE 123-bus system is a three-phase, unbalanced, radial distribution system with single-, double-, or three-phase buses, and it has $263$ phases in total. The data we use for our implementation were simulated at 1-minute resolution using power flow analysis with diversified load and solar profiles created for each bus using real solar irradiance and load consumption data. 
We first build our tensor using $72$ consecutive time step (i.e., $72$ minutes) data during the time slots with high photovoltaic generation. In addition, we assess the performance of the proposed tensor decomposition approach on a 1-day data set comprising $72$ samples, i.e., we sample the data every $20$ minutes. These data have much higher variability than the consecutive data. 
 
 The tensor is of size $263\times 5\times 72$ for both cases, where $263$ is the number of phases in the considered multiphase network. For consecutive data, the voltage magnitudes range from $0.9131$ to $1$ p.u. with mean $0.9575$, the nonzero active powers range from $-0.10414$ to $13.1941$ KW with a mean of $0.6394$, and the nonzero reactive powers range from $-0.5207$ to $7.4006$ KW with a mean of $0.3339$. 
 
 We use the mean absolute percentage error (MAPE) for voltage magnitude ($|V|$) and mean absolute error (MAE) for voltage angle ($\theta$), active power ($P$), and reactive power ($Q$) to measure the performance of the  proposed approach~\eqref{eq:impute-opt}. The MAPE and MAE are defined as:
$\text{MAPE}: = \frac{1}{n}\sum\frac{\text{Actual-Forecast}}{\text{Actual}}\times 100\%$ and $\text{MAE}: = \frac{1}{n}\sum(\text{Actual-Forecast})$, where $n$ is the total number of elements. When we calculate the MAEs for active power and reactive power, we consider only the nonzero load phases. The active power and reactive power are considered to be known as $0$ for zero load phases. The simulation results in Section~\ref{subsec:slabsamplingsimulation} and Section \ref{subsec:fibersamplingsimulation} are based on the average of $50$ runs of each scenario.

 \subsection{Low-Rank Property}
The tensors comprising consecutive and nonconsecutive data are both of size $263\times 5\times 72$. Using the identifiability condition~\eqref{eq:indetfiability}, if the rank ($F$) of the tensors satisfies $F\leq 2^{\floor{5}+\floor{72}-2}=32$ and the incomplete tensors satisfy the sampling conditions in Section~\ref{subsec:slabsampling} and Section \ref{subsec:fibersampling}, respectively, then the optimization formulation~\eqref{eq:impute-opt} can recover the ground-truth factors of the full tensor almost surely. 

\begin{figure}
    \centering
  \includegraphics[scale=0.5,trim= 60 200 20 200, clip]{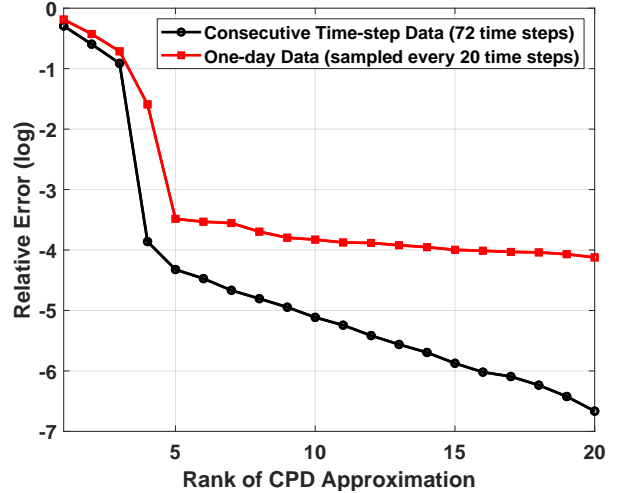}
    \caption{Relative error between the data and the computed CPD approximation}
    \label{fig:lowrank}
\end{figure}

First, we assess that the tensor $\Ten{X}$ is of low rank for both the consecutive and nonconsecutive time instances. We compute the relative errors between the tensors and their rank-$k$ ($k=1,\ldots, 20$) CPD approximations, which are depicted in Fig.~\ref{fig:lowrank}. The relative error between tensor $\Ten{X}\in\mathbb{R}^{I\times J\times K}$ and its rank-$k$ CPD approximation $\Tilde{\Ten{X}}_k\in\mathbb{R}^{I\times J\times K}$ is defined as the ratio of the Frobenius norm square of the residual and the Frobenius norm square of the tensor, i.e., $||\Ten{X}-\Tilde{\Ten{X}}_k||_\text{F}^2/||\Ten{X}||_\text{F}^2$, where $||\Ten{X}||_\text{F}^2:=\sum_{i=1}^I\sum_{j=1}^J\sum_{k=1}^K\Ten{X}(i,j,k)^2$. This can be understood as the variance of $\Ten{X}$, which is captured by the $k$-th rank approximation $\Tilde{\Ten{X}}_k$.

Fig.~\ref{fig:lowrank} shows that the relative error decreases as the rank of the CPD approximation increases for both tensors using consecutive data and nonconsecutive data. For the consecutive case, the relative error achieves $10^{-4}$ for rank-$5$ approximation, and it decreases obviously as the rank increases. For the nonconsecutive case, the relative error decreases to $10^{-4}$ for a relatively low rank; hence, we conclude that the considered tensors can be well approximated using low-rank approximations.

\subsection{Slab Sampling}
\label{subsec:slabsamplingsimulation}
As shown in the previous subsection, a low-rank CPD model is effective in describing the state tensor; hence, we assess the performance of the proposed state imputation approach with a CPD model of rank $11$.
To guarantee that the incomplete tensor can be recovered almost surely, using the conditions in Proposition~\ref{prop:slabsampling}, the number of sampled horizontal slabs ($I_h$) and the number of sampled frontal slabs ($K_f$) should satisfy: $I_h\geq 16$, and $K_f\geq 3$. Having $I_h = 16$ means that all measurements for the sampled $16$ phases out of $263$ phases are known for every time instance, and $K_f = 3$ means that all measurements for all phases are known at the sampled $3$ time steps. We will consider that $16$ phases and $3$ time steps are sampled in our simulation.  The $16$ phases sampled include the $3$ slack bus phases in addition to another $13$ phases chosen randomly from the nonzero load phases. The $3$ sampled time steps are time steps $24$, $48$, and $72$. Including the zero active power and reactive power known, the available measurements' percentage is $34.05\%$.
We solve the optimization problem~\eqref{eq:impute-opt} using the  \textsc{sdf-nls} solver to recover the factors of the state tensor: $\hat{\A}, {\hat{\B}}, \hat{\C}$; hence, we reconstruct the estimate tensor $\hat{\Ten{X}} := \cpd{\hat{\A}}{\hat{\B}}{\hat{\C}}$. To demonstrate the efficacy of our approach, we also assess its performance on the consecutive and nonconsecutive data with $1\%$ Gaussian noise added to the observed measurements.

The performance results in terms of MAPE for voltage magnitude and MAEs for voltage angle, active power, and reactive power for the noiseless consecutive data and the noisy consecutive data with $1\%$ Gaussian noise under the condition of $I_h = 16, K_f = 3$ are  shown in Table~\ref{tab:slabsampling_consecutive}; and the corresponding results for the noiseless nonconsecutive data and the noisy nonconsecutive data with $1\%$ Gaussian noise are shown in Table~\ref{tab:slabsampling_nonconsecutive}.

Table~\ref{tab:slabsampling_consecutive} shows the efficacy of the approach in terms of the MAPE for voltage magnitude and MAEs for voltage angle, active power, and reactive power. Even though the MAPE and MAEs are much bigger for the noisy case, they are small enough compared to the noise level, which can be seen by comparing the last two rows of Table~\ref{tab:slabsampling_consecutive}. This means that the proposed CPD model also denoises the noisy measurements to a certain extent.

\begin{figure}
    \centering
  \includegraphics[scale=0.5,trim= 45 230 50 250, clip]{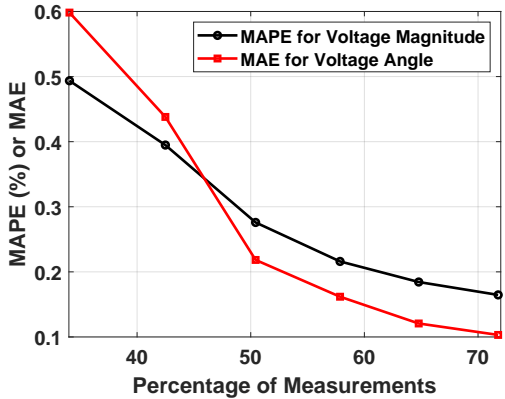}
    \caption{MAPE ($|V|$) and MAE ($\theta$) vs. percentage of measurements for noiseless nonconsecutive data}
    \label{fig:slabsampling1}
\end{figure}

\begin{figure}
    \centering
  \includegraphics[scale=0.5,trim= 50 240 50 250, clip]{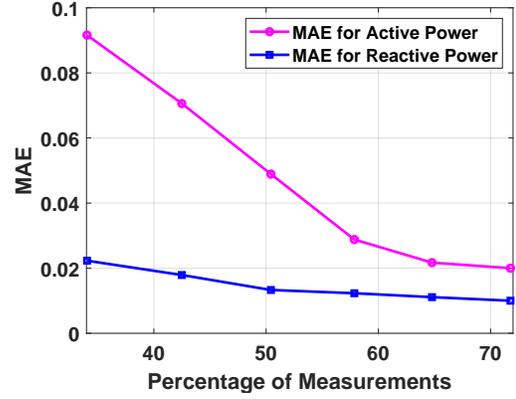}
    \caption{MAE ($P$) and MAE ($Q$) vs. percentage of measurements for noiseless nonconsecutive data}
    \label{fig:slabsampling2}
\end{figure}

Table~\ref{tab:slabsampling_nonconsecutive} shows that the MAPE and MAEs for the noiseless nonconsecutive data are much bigger than those for the consecutive data. This can be explained by Fig.~\ref{fig:lowrank}, which shows that the low-rank model is able to achieve better performance for the consecutive data scenario.
As more horizontal and frontal slabs are sampled, however, the MAPE and MAEs decrease dramatically, which are shown in Figs.~\ref{fig:slabsampling1} and \ref{fig:slabsampling2}. In Figs.~\ref{fig:slabsampling1} and \ref{fig:slabsampling2}, we use the $16$ phases and $3$ time steps sampled as the baseline, and then we consider the following cases: $32$ phases + $6$ time steps, $48$ phases + $9$ time steps, $64$ phases + $12$ time steps, $80$ phases + $15$ time steps, and $96$ phases + $18$ time steps. The phases sampled are selected randomly from nonzero load phases, and the sampled time steps are always chosen to be equally spaced along the third dimension of the tensor. The horizontal axis represents the measurement percentage for each case. 

\begin{table}
	\renewcommand{\arraystretch}{1.2}
	\caption{Performance of model~(\ref{eq:impute-opt}) using rank-11 approximation for consecutive data}
\label{tab:slabsampling_consecutive}
	\begin{center}
		\begin{tabular}{l c c c c} 
			\toprule
	            & MAPE($|V|$)  & MAE($\theta$) & MAE($P$) & MAE($Q$)\\
			\midrule
			Noiseless &  $0.0792\%$ & $0.0338$ & $0.0119$ & $ 0.0060$\\
			1\% Noise & $0.5485\%$ & $0.2650 $ & $0.0140$ & $0.0082$\\
			Data vs Noisy Data &  $0.8313\%$ & $0.4756$ & $0.0079$ & $0.0078$\\
			\bottomrule
		\end{tabular}
	\end{center}
\end{table}

\begin{table}
	\renewcommand{\arraystretch}{1.2}
	\caption{Performance of model~(\ref{eq:impute-opt}) using rank-11 approximation for nonconsecutive data}
\label{tab:slabsampling_nonconsecutive}
	\begin{center}
		\begin{tabular}{l c c c c} 
			\toprule
	            & MAPE($|V|$)  & MAE($\theta$) & MAE($P$) & MAE($Q$)\\
			\midrule
			Noiseless &  $0.4936\%$ & $0.5983$ & $0.0916$ & $ 0.0223$\\
			1\% Noise &$0.7391\%$ & $0.6933$ & $0.0922$ & $0.0227$\\
			\bottomrule
		\end{tabular}
	\end{center}
\end{table}

\subsection{Fiber Sampling}
\label{subsec:fibersamplingsimulation}
In this subsection, we use optimization problem~\eqref{eq:impute-opt} with rank $F = 8$ approximation and $\Ten{M} = \Ten{M}_f$ to recover the latent factors of the state tensor $\Ten{X}$; hence, we reconstruct a tensor $\Ten{X}$ containing estimates of the unobserved values in $\Ten{X}$. 
We consider a fiber-sampling strategy where two fiber patterns are sampled. The first pattern samples the real voltage, imaginary voltage, and voltage magnitude for some phases at all time steps, and the second one samples the active power and reactive power for some phases at all time steps. We assume that only the three slack bus phases are sampled for both patterns. Therefore, we have: $\mathcal{S}_r^{(1)}\cap \mathcal{S}_r^{(2)}=\{1,2,3\}$, $\mathcal{S}_c^{(1)}=\{1,2,3\}$, and $\mathcal{S}_c^{(2)}=\{4,5\}$,  which satisfies conditions 1), 3), and 4) in Proposition~\ref{prop:fibersampling}. The conditions 2) and 5) in Proposition~\ref{prop:fibersampling} require that $|\mathcal{S}_r^{(1)}|+ |\mathcal{S}_r^{(2)}|=266$ and  $|\mathcal{S}_r^{(1)}|, |\mathcal{S}_r^{(2)}|\geq 16$. We implement model~(\ref{eq:impute-opt}) on the consecutive and nonconsecutive data under the assumption of $|\mathcal{S}_r^{(1)}|=250$ and $|\mathcal{S}_r^{(2)}|=16$. The $16$ phases sampled where the active power and reactive power are sampled at all time instances include the $3$ phases at the slack bus and $13$ randomly selected phases from the nonzero load phases. The $250$ phases sampled with the real voltage, imaginary voltage, and voltage magnitude at all time steps include the $3$ slack bus phases and the $247$ phases left.
\begin{table}
	\renewcommand{\arraystretch}{1.2}
	\caption{Performance of model~(\ref{eq:impute-opt}) using rank-8 approximation for consecutive data}
\label{tab:fibersampling_consecutive}
	\begin{center}
		\begin{tabular}{l c c c c} 
			\toprule
	            & MAPE($|V|$)  & MAE($\theta$) & MAE($P$) & MAE($Q$)\\
			\midrule
			Noiseless &  $0.2242\%$ & $0.0961$ & $0.0150$ & $ 0.0080$\\
			1\% Noise &$0.4635\%$  & $0.2082$  &$0.0340$  & $0.0200$\\
			\bottomrule
		\end{tabular}
	\end{center}
\end{table}

\begin{table}
	\renewcommand{\arraystretch}{1.2}
	\caption{Performance of model~(\ref{eq:impute-opt}) using rank-8 approximation for nonconsecutive data}
\label{tab:fibersampling_nonconsecutive}
	\begin{center}
		\begin{tabular}{l c c c c} 
			\toprule
	            & MAPE($|V|$)  & MAE($\theta$) & MAE($P$) & MAE($Q$)\\
			\midrule
			Noiseless &  $0.2569\%$ & $0.1151$ & $0.1101$ & $ 0.0405$\\
			1\% Noise &$0.4800\%$  & $0.2326$  &$0.1345$  & $0.0741$\\
			\bottomrule
		\end{tabular}
	\end{center}
\end{table}

Table~\ref{tab:fibersampling_consecutive} shows the performance of the proposed approach with rank $F=8$ approximation in case of consecutive data. The results show that the proposed approach achieves acceptable MAPE for voltage magnitude and MAEs for voltage angle, active power, and reactive power under both noiseless and noisy cases. The results achieved under noisy data are less accurate, which is expected because of the noise effect; however, the voltage estimation results are still less than $0.5\%$ for MAPE, which is significantly less than the often required estimation accuracy of $1\%$. This is also shown in Table~\ref{tab:fibersampling_nonconsecutive}, which presents the results for the nonconsecutive data case. Comparing Table~\ref{tab:fibersampling_consecutive} and Table~\ref{tab:fibersampling_nonconsecutive}, the results demonstrate the efficacy of the proposed approach in estimating the state of the network from samples. For the slab sampling, however, the performance for the noiseless consecutive case is much better than the noiseless nonconsecutive data, which can be seen by comparing Table~\ref{tab:slabsampling_consecutive} and Table~\ref{tab:slabsampling_nonconsecutive}. This can be explained because the data variability for the consecutive data is less than that of the nonconsecutive data. Although all phases at all time steps are sampled with either pattern of fiber sampling, no full slab information is known. Because the sampled fibers collect information regarding all time instances, the performance of the approach is not affected by the huge variability in the third dimension.

\begin{figure}
    \centering
  \includegraphics[scale=0.5,trim= 50 240 50 250, clip]{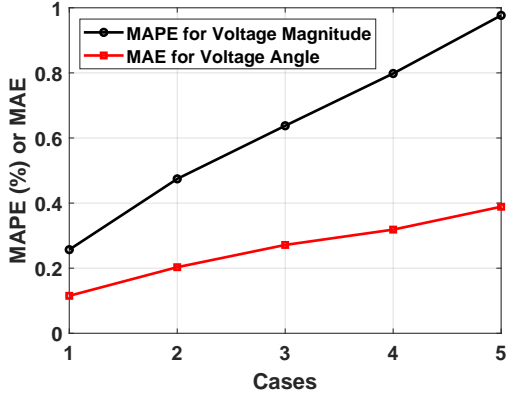}
    \caption{MAPE ($|V|$) and MAE ($\theta$) vs. cases for noiseless nonconsecutive data}
    \label{fig:fibersampling1}
\end{figure}

\begin{figure}
    \centering
  \includegraphics[scale=0.5,trim= 50 240 50 250, clip]{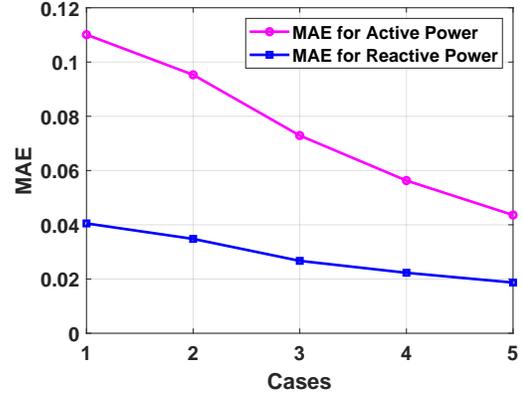}
    \caption{MAE ($P$) and MAE ($Q$) vs. cases for noiseless nonconsecutive data}
    \label{fig:fibersampling2}
\end{figure}

For the nonconsecutive data case, as $|\mathcal{S}_r^{(1)}|$ increases and $|\mathcal{S}_r^{(2)}|$ decreases when they still satisfy the conditions in Proposition~\ref{prop:fibersampling}, the MAPE for voltage magnitude and MAE for voltage angle increase, whereas the MAEs for active power and reactive power decrease, as shown in Figs.~\ref{fig:fibersampling1} and \ref{fig:fibersampling2}. This is because increasing $ |\mathcal{S}_r^{(1)}|$ increases the observability of the power quantities while decreasing $|\mathcal{S}_r^{(2)}|$ decreasing the observed voltage quantities.  Case 1 is the case of $|\mathcal{S}_r^{(1)}|=16$ and $|\mathcal{S}_r^{(2)}|=250$. Cases 2--5 are: $|\mathcal{S}_r^{(1)}|=32$, $|\mathcal{S}_r^{(2)}|=234$; $|\mathcal{S}_r^{(1)}|=48$, $|\mathcal{S}_r^{(2)}|=218$; $|\mathcal{S}_r^{(1)}|=64$, $|\mathcal{S}_r^{(2)}|=202$; and $|\mathcal{S}_r^{(1)}|=80$ and $|\mathcal{S}_r^{(2)}|=186$, respectively. 

\subsection{Comparison of Model-Nased Matrix Completion and Tensor Completion}
The references \cite{donti2019matrix} and \cite{Sagan2019} both applied the conventional low-rank matrix completion model with power flow constraints (model-based matrix completion) for state imputation. In \cite{donti2019matrix}, the data matrix was built using a single snapshot of data, whereas \cite{Sagan2019} built the data matrix using multiple time-step data and developed a decentralized method using a proximal alternating direction method of multipliers. We compare our proposed model-free tensor completion approach to the model-based matrix completion  method in \cite{Sagan2019} using the same noiseless consecutive and nonconsecutive data under the sampling setting of $16$ horizontal slabs and $3$ frontal slabs. The comparison results in terms of MAPE for voltage magnitude and MAEs for voltage angle, active power, and the reactive power are shown
in Table~\ref{tab:comparisonofmcandslabopt} and Table \ref{tab:comparisonofmcandslaboptnoncont} for the consecutive and nonconsecutive data cases, respectively. For the consecutive case, our proposed approach achieves better voltage magnitude estimation, whereas the MAEs for the other quantities are of the same order.
For the nonconsecutive case, the results for the matrix completion are generally better than our model but of the same order.
Note that our proposed approach achieves this result even without the use of the model information required by the matrix completion method~\cite{Sagan2019}. 

\begin{table}
	\renewcommand{\arraystretch}{1.2}
	\caption{Comparison of model-based matrix completion and model~(\ref{eq:impute-opt}) for noiseless consecutive data}
\label{tab:comparisonofmcandslabopt}
	\begin{center}
		\begin{tabular}{l c c c c} 
			\toprule
	            & MAPE($|V|$)  & MAE($\theta$) & MAE($P$) & MAE($Q$)\\
			\midrule
			Tensor Sampling &  $0.0792\%$ & $0.0338$ & $0.0119$ & $ 0.0060$\\
			Matrix Completion~\cite{Sagan2019} & $0.3023\%$ & $0.0419 $ & $0.0076$ & $0.0026$\\
			\bottomrule
		\end{tabular}
	\end{center}
\end{table}

\begin{table}
	\renewcommand{\arraystretch}{1.2}
	\caption{Comparison of model-based matrix completion and model~(\ref{eq:impute-opt}) for noiseless nonconsecutive data}
\label{tab:comparisonofmcandslaboptnoncont}
	\begin{center}
		\begin{tabular}{l c c c c} 
			\toprule
	            & MAPE($|V|$)  & MAE($\theta$) & MAE($P$) & MAE($Q$)\\
			\midrule
			Tensor Sampling &  $0.4936\%$ & $0.5983$ & $0.0916$ & $ 0.0223$\\
			Matrix Completion~\cite{Sagan2019} & $0.4312\%$ & $0.2965 $ & $0.0181$ & $0.0041$\\
			\bottomrule
		\end{tabular}
	\end{center}
\end{table}


\section{Conclusions and Future Research}
This paper presented a novel formulation of the state estimation problem in distribution networks using tensor decomposition. By stacking the network states at multiple time instances, we formed a three-way tensor that enjoys a low-rank structure. This structure was used to formulate the state estimation as a tensor imputation problem. Two principle sampling schemes were considered: slab and fiber sampling. For both schemes, this paper presented sufficient sampling requirements that guarantee recoverability of the latent factors of the full state tensor. In addition, simulations were carried out on semi-real data for the IEEE $123$-bus distribution feeder, where both sampling schemes showed high estimation accuracy in several scenarios.

The current version of this approach does not allow for the incorporation of available model information. An extension of this framework that allows for fusing model information with measurements using constrained tensor factorization methods. In addition, distributed implementation for the proposed approach will be sought to allow for scalable decentralized implementation on advanced metering infrastructure.

\section*{Acknowledgment}

The authors would thank Charilaos Kanatsoulis for the insightful discussions at early stages of this work. The authors acknowledge the support of the work by the Laboratory Directed Research and Development Program at the National
Renewable Energy Laboratory, Golden, CO.


%





\ifCLASSOPTIONcaptionsoff
  \newpage
\fi





\bibliographystyle{IEEEtran}
\bibliography{IEEEabrv,Bibliography,references}

\vfill


\end{document}